\documentclass[aps,pra,floatfix,nofootinbib,groupedaddress,superscriptaddress,citesort]{revtex4}
\usepackage[T1]{fontenc}
\usepackage[utf8]{inputenc}
\usepackage{mathrsfs}
\usepackage{amsfonts}
\usepackage{amstext}
\usepackage{amsmath}
\usepackage{amssymb}
\usepackage[dvipsnames]{xcolor}
\usepackage[dvips]{graphicx}
\usepackage{color}
\def\qed{\leavevmode\unskip\penalty9999 \hbox{}\nobreak\hfill
     \quad\hbox{\leavevmode  \hbox to.77778em{%
              \hfil\vrule   \vbox to.675em%
               {\hrule width.6em\vfil\hrule}\vrule\hfil}}
     \par\vskip3pt}

\newtheorem{theorem}{Theorem}

\begin{document}

\title{Geometric discord for multiqubit systems}
\author{Chen-Lu Zhu}
\affiliation{Department of Mathematics, East China University of Technology, Nanchang 330013, China}
\author{Bin Hu\footnote{Corresponding author}}
 \email{bhu@ecut.edu.cn.}
\affiliation{Department of Mathematics, East China University of Technology, Nanchang 330013, China}

\author{Bo Li}
\email{libobeijing2008@163.com.}
\affiliation{School of Mathematics and Computer science, Shangrao Normal University, Shangrao 334001, China}
\author{Zhi-Xi Wang}
\affiliation{School of Mathematical Sciences, Capital Normal University, Beijing 100048, China}
\author{Shao-Ming Fei}
\affiliation{School of Mathematical Sciences, Capital Normal University, Beijing 100048, China}
\affiliation{Max-Planck-Institute for Mathematics in the Sciences, 04103, Leipzig, Germany}

\begin{abstract}
Radhakrishnan \emph{et.al} [Phys. Rev. Lett. \textbf{124}, 110401 (2020)]  proposed quantum discord to multipartite systems and derived  explicit  formulae for any states.
These results are significant in capturing quantum correlations for multi-qubit systems.
In this paper, we evaluate the geometric measure of multipartite quantum discord and obtain results for a large family of multi-qubit states.
Furthermore, we investigated the dynamic behavior of geometric discord for the family of two-, three- and four-qubit states under phase noise acting on the first qubit.
And we discover that sudden change of multipartite geometric discord can appear when phase noise act only on one part of the two-, three- and four-qubit states.
\end{abstract}

\maketitle
\parskip=4pt
\section{\bf Introduction}
In the early research of quantum information, entanglement was considered to be an important resource, which was used to distinguish the quantum world from the classical world.
Compared with traditional computing, quantum computing was considered to have tremendous advantages via exploiting entanglement, otherwise, it would lose its competitive superiority.
For a long time, people focused on the research of quantum information on quantum entanglement, believing that "entanglement is not only one of many characteristics, but the characteristic of quantum physics".
However, with the development of research, it is found that entanglement is only a subset of quantum correlations, and many quantum states without entanglement can still exhibit their quantum properties in quantum information processing.
Ollivier and Zurek \cite{Ollivier} and Henderson and Vedral \cite{Henderson} introduced a measure called quantum discord, which captures not only the quantum correlations of entangled states but also the separable states. Over the next two decades, it has received a lot of attention \cite{Rulli,Sone,Chanda,Rana,Luo3,Luo4,Girolami,Hunt,Lang,Li2,Dakic2,Modi}.

For bipartite systems, quantum discord is defined as the difference between two natural quantum extensions of the classical mutual information.
In some special cases, the analytical results of bipartite quantum discord are known\cite{Luo4,Li2}.
Recently, Radhakrishnan \emph{et.al} \cite{Radhakrishnan} proposed a definition of multipartite quantum discord which is in
consistent with the original bipartite definition \cite{Ollivier,Henderson}.
In \cite{Li} we considered the following family of $N$-qubit states,
\begin{equation}\label{rho}
\rho=\frac{1}{2^N}(I+\sum\limits_{j=1}^3c_j\sigma_j\otimes\cdots\otimes\sigma_j),
\end{equation}
where $I$ is the identity operator, $c_j$ are real constants satisfying certain constraints and $\sigma_j$, $j=1,2,3$, are the  Pauli matrices.
We derived analytical formulae for quantum discord of $(2v+1)$, $(4v-2)$ and $(4v)$-qubit states.
In general, it is difficult to evaluate quantum discord due to the complexity of the optimization.
For this reason, Daki{\'c} \emph{et.al} \cite{Dakic} introduced the following geometric measure of quantum discord for bipartite states:
\begin{equation}\label{dg}
D^{(2)}_G(\rho):=\min_{ \chi\in\Omega}||\rho-\chi||^2,
\end{equation}
where $\Omega$ denotes the set of zero-discord states and the geometric quantity $||\rho-\chi||^2={\rm Tr}(\rho-\chi)^2$ is the square of Hilbert-Schmidt norm of Hermitian operators. The geometric measure of quantum discord has attracted much attention, mainly because of its computational simplicity \cite{Fel'dman,Paula,Costa,Zhou,Bellomo,Debarba,Brown,Miranowicz,Tufarelli,Rana2,Hassan,Passante}.
In particular, Luo and Fu \cite{Luo2} evaluated the geometric measure of quantum discord and obtained explicit tight lower bounds for arbitrary states.
However, Piani \cite{Piani} argued that the geometric discord may not be a good measure for the quantumness of correlations,
since it may increase even under trivial local reversible operations.
Subsequently, Chang and Luo \cite{Chang} showed that this geometric discord problem can be remedied simply by starting from the square root of a density operator, rather than the density operator itself.
Nevertheless, the generalizations of geometric discord to tripartite and multipartite systems remain open.

On the other hand, Maziero \emph{et.al} \cite{Maziero} studied the dynamical behavior of quantum discord under decoherence.
Later Jia \emph{et.al} \cite{Jia} discovered that even when part of the composite entangled state is exposed to a noisy environment, the quantum correlation changes suddenly.

In this article, we propose the concept of geometric measure of multipartite quantum discord and
evaluate its value for the family of $N$-qubit states given in (\ref{rho}).
We investigate the dynamics of geometric discord for the family of two-, three- and four-qubit states with only one qubit being exposed to noise.
The article is organized as follows. In Sec. \uppercase\expandafter{\romannumeral2}, we calculate analytically the multi-qubit geometric discord for a family of quantum states.
In Sec. \uppercase\expandafter{\romannumeral3} we show that sudden change of geometric discord for multipartite system can occur when the phase noise acts only on one qubit of
the family of two-, three- and four-qubit states.
Finally, Sec. \uppercase\expandafter{\romannumeral4} is devoted to conclusion.

\section{Geometric discord for multi-qubit systems}
In the definition of bipartite geometric discord (\ref{dg}), only one of the subsystems is measured.
This is sufficient because the correlations are only between two subsystems for bipartite cases.
For $N$-partite systems, $N-1$ local measurements are needed to measure all the quantum correlations \cite{Rulli},
where each measurement depends conditionally on the previous measurement outcomes.
The $(N-1)$-partite measurement is given by
\begin{eqnarray}
\Pi_{j_1\cdots j_{N-1}}^{A_1\cdots A_{N-1}}=\Pi_{j_1}^{A_1}\otimes\Pi_{j_2|j_1}^{A_2}\cdots \otimes \Pi_{j_{N-1}|j_1\cdots j_{N-2}}^{A_{N-1}},\nonumber
\end{eqnarray}
where $A_i$ labels the $N$ subsystems, $\Pi_{j_1}^{A_1}$ is a von Neumann projection operator on the subsystem $A_1$,
$\Pi_{j_1|j_2}^{A_2}$ is a projector on subsystem $A_2$, conditioned on the measurement outcome on $A_1$.
The measurements are given in the following order: $A_1 \rightarrow A_2 \rightarrow \cdots  \rightarrow A_{N-1}.$
We define the following geometric discord for multi-qubit systems,
\begin{equation}\label{ndg}
D_G^{(N)}(\rho):=\min_{ \chi\in\Omega}||\rho-\chi||^2,
\end{equation}
where the distance $||\rho-\chi||$ between states $\rho$ and $\chi$ is given by
\begin{eqnarray}\label{rhochi}
||\rho-\chi||^2=||\rho||^2-2{\rm Tr}\rho\chi+||\chi||^2.
\end{eqnarray}

Consider the family of $N$-qubit states given in (\ref{rho}), which reduce to the well-known Bell-diagonal states for $N=2$.
And the geometric measure of its quantum discord has been shown in \cite{Yao}, which is
\begin{eqnarray}\label{dg2}
D^{(2)}_G(\rho)=\frac{1}{4}(c_1^2+c_2^2+c_3^2-max\{c_1^2,c_2^2,c_3^2\}).
\end{eqnarray}

For the case of $N=3$, the states associated with the subsystems $A$, $B$, and $C$ are given as
\begin{equation}\label{rho3}
\rho=\frac{1}{8}(I+\sum\limits_{j=1}^3c_j\sigma_j\otimes\sigma_j\otimes\sigma_j).
\end{equation}
To evaluate the tripartite geometric discord $D^{(3)}_G(\rho)$ defined in (\ref{ndg}), one needs to
to calculate $||\rho||^2$, $-2{\rm Tr}\rho\chi$ and $||\chi||^2$ according to (\ref{rhochi}).
We have
\begin{equation}\label{rho^2}
||\rho||^2={\rm Tr}(\rho^2)=\frac{1}{8}(1+c_1^2+c_2^2+c_3^2).
\end{equation}

To evaluate $-2{\rm Tr}\rho\chi$ and $||\chi||^2$, we need to measure the subsystem $A$.
Let $\{\Pi_k=|k\rangle\langle k|:k=0,1\}$,
any von Neumann measurement on subsystem $A$ is given by
$\{A_k=V_A\Pi_kV_A^\dagger:k=0,1\}$, where $V_A=t_AI+i\vec{y}_A\vec{\sigma}$ is the unitary operator with $t_A\in \mathbb{R}$, $\vec{y}_A=(y_{A1},y_{A2},y_{A3})\in \mathbb{R}^3$ and $t_A^2+y_{A1}^2+y_{A2}^2+y_{A3}^2=1$.
After the measurement on ${A_k}$, the state $\rho$ is going to become the ensemble $\{\rho_k,p_k\}$, where $\rho_k=\frac{1}{p_k}(A_k\otimes I)\rho(A_k\otimes I)$
and $p_k={\rm Tr}(A_k\otimes I)\rho(A_k\otimes I).$ We obtain
$p_0=p_1=\frac{1}{2}$, and
\begin{eqnarray}
\rho_0=\frac{1}{4}V_{A}\Pi_0V_A^\dagger\otimes(I+c_1d_{1}\sigma_1\otimes\sigma_1+c_2d_{2}\sigma_2\otimes\sigma_2+c_3d_{3}\sigma_3\otimes\sigma_3),
\label{rho0}
\end{eqnarray}
\begin{eqnarray}
\rho_1=\frac{1}{4}V_{A}\Pi_1V_A^\dagger\otimes(I-c_1d_{1}\sigma_1\otimes\sigma_1-c_2d_{2}\sigma_2\otimes\sigma_2-c_3d_{3}\sigma_3\otimes\sigma_3),
\label{rho1}
\end{eqnarray}
where
\begin{eqnarray}
& d_{1}=2(-t_Ay_{A2}+y_{A1}y_{A3}),\nonumber \\
& d_{2}=2(t_Ay_{A1}+y_{A2}y_{A3}),\nonumber \\
& d_{3}=t_A^2-y_{A1}^2-y_{A2}^2+y_{A3}^2.\nonumber
\end{eqnarray}

Next, we consider the subsystem $B$ according to the measurement results from $A$.
With respect to the outcome $l$ ($l=0,1$) of the measurement on $A$, we denote $\{B_k^l=V_{B^l}\Pi_kV_{B^l}^\dagger:k=0,1\},~l=0,1,$
be the local measurement on the subsystem $B$ when the outcome of measurement on $A$ is $j(j=0,1)$, where the unitary $V_{B^l}=t_{B^l}I+i\vec{y}_{B^l}\vec{\sigma}$ with $t_{B^l}\in \mathbb{R}$, $\vec{y}_{B^l}=(y_{B^l1},y_{B^l2},y_{B^l3})\in \mathbb{R}^3$ and $t_{B^l}^2+y_{B^l1}^2+y_{B^l2}^2+y_{B^l3}^2=1$.

Note that the subsystems $B$ and $C$ in $\rho_{0}$ are still in a Bell-diagonal state.
Applying the measurement $\{B_k^0:k=0,1\}$, we have
 \begin{eqnarray}
\rho_{00}&=&\frac{1}{2}V_A\Pi_0V_A^\dagger\otimes V_{B^0}\Pi_0V_{B^0}^\dagger\otimes(I+c_1d_{1}e_1\sigma_1+c_2d_{2}e_2\sigma_2+c_3d_{3}e_3\sigma_3),\nonumber
\label{rho00}
\end{eqnarray}
\begin{eqnarray}
\rho_{01}&=&\frac{1}{2}V_A\Pi_0V_A^\dagger\otimes V_{B^0}\Pi_1V_{B^0}^\dagger\otimes(I-c_1d_{1}e_1\sigma_1-c_2d_{2}e_2\sigma_2-c_3d_{3}e_3\sigma_3),\nonumber
\label{rho01}
\end{eqnarray}
where
\begin{eqnarray}
& e_{1}=2(-t_{B^0}y_{B^02}+y_{B^01}y_{B^03}),\nonumber \\
& e_{2}=2(t_{B^0}y_{B^01}+y_{B^02}y_{B^03}),\nonumber\\
& e_{3}=t_{B^0}^2-y_{B^01}^2-y_{B^02}^2+y_{B^03}^2.\nonumber
\end{eqnarray}
For the state $\rho_1$, after measuring the subsystem $B$ one has
 \begin{eqnarray}
\rho_{10}&=&\frac{1}{2}V_A\Pi_1V_A^\dagger\otimes V_{B^1}\Pi_0V_{B^1}^\dagger\otimes(I-c_1d_{1}f_1\sigma_1-c_2d_{2}f_2\sigma_2-c_3d_{3}f_3\sigma_3),\nonumber
\label{rho10}
\end{eqnarray}
\begin{eqnarray}
\rho_{11}&=&\frac{1}{2}V_A\Pi_1V_A^\dagger\otimes V_{B^1}\Pi_1V_{B^1}^\dagger\otimes(I+c_1d_{1}f_1\sigma_1+c_2d_{2}f_2\sigma_2+c_3d_{3}f_3\sigma_3),\nonumber
\label{rho10}
\end{eqnarray}
where
\begin{eqnarray}
& f_{1}=2(-t_{B^1}y_{B^12}+y_{B^11}y_{B^13}),\nonumber \\
& f_{2}=2(t_{B^1}y_{B^11}+y_{B^12}y_{B^13}),\nonumber \\
& f_{3}=t_{B^1}^2-y_{B^11}^2-y_{B^12}^2+y_{B^13}^2.\nonumber
\end{eqnarray}

The state $\chi$ is given as $\chi=p_{00}\rho_{00}+p_{01}\rho_{01}+p_{10}\rho_{10}+p_{11}\rho_{11}$. Then
\begin{eqnarray}\label{rhochi2}
-2{\rm Tr}(\rho\chi)=-\frac{1}{4}[{\rm Tr}(I\chi)+{\rm Tr}(\sum\limits_{j=1}^3c_j\sigma_j\otimes\sigma_j\otimes\sigma_j\chi)]\nonumber\\
=-\frac{1}{4}[1+\frac{1}{2}(c_1^2d_1^2e_1^2+c_1^2d_1^2f_1^2+c_2^2d_2^2e_2^2+c_2^2d_2^2f_2^2+c_3^2d_3^2e_3^2+c_3^2d_3^2f_3^2)].
\end{eqnarray}
Let $q$ be the quantum states of the system $C$,
\begin{eqnarray}
q_{00}=I+c_1d_{1}e_1\sigma_1+c_2d_{2}e_2\sigma_2+c_3d_{3}e_3\sigma_3,\nonumber\\
q_{01}=I-c_1d_{1}e_1\sigma_1-c_2d_{2}e_2\sigma_2-c_3d_{3}e_3\sigma_3,\nonumber\\
q_{10}=I-c_1d_{1}f_1\sigma_1-c_2d_{2}f_2\sigma_2-c_3d_{3}f_3\sigma_3,\nonumber\\
q_{11}=I+c_1d_{1}f_1\sigma_1+c_2d_{2}f_2\sigma_2+c_3d_{3}f_3\sigma_3.\nonumber
\end{eqnarray}
One can verify that
\begin{eqnarray}
{\rm Tr}(q_{00}^2)={\rm Tr}(q_{01}^2)=2(1+c_1^2d_1^2e_1^2+c_2^2d_2^2e_2^2+c_3^2d_3^2e_3^2),\nonumber\\
{\rm Tr}(q_{10}^2)={\rm Tr}(q_{11}^2)=2(1+c_1^2d_1^2f_1^2+c_2^2d_2^2f_2^2+c_3^2d_3^2f_3^2).\nonumber
\end{eqnarray}
Hence,
\begin{eqnarray}\label{chi2}
{\rm Tr}(\chi^2)&=&\frac{1}{8^2}{\rm Tr}(V_{A}\Pi_0V_{A}^\dagger)^2{\rm Tr}(V_{{B}^0}\Pi_0V_{{B}^0}^\dagger)^2{\rm Tr}(q_{00}^2)\nonumber\\
&&+\frac{1}{8^2}{\rm Tr}(V_{A}\Pi_0V_{A}^\dagger)^2{\rm Tr}(V_{{B}^0}\Pi_1V_{{B}^0}^\dagger)^2{\rm Tr}(q_{01}^2)\nonumber\\
&&+\frac{1}{8^2}{\rm Tr}(V_{A}\Pi_1V_{A}^\dagger)^2{\rm Tr}(V_{{B}^1}\Pi_0V_{{B}^1}^\dagger)^2{\rm Tr}(q_{10}^2)\nonumber\\
&&+\frac{1}{8^2}{\rm Tr}(V_{A}\Pi_1V_{A}^\dagger)^2{\rm Tr}(V_{{B}^1}\Pi_1V_{{B}^1}^\dagger)^2{\rm Tr}(q_{11}^2)\nonumber\\
&=&\frac{1}{16}(2\!+\!c_1^2d_1^2e_1^2\!+\!c_1^2d_1^2f_1^2\!+\!c_2^2d_2^2e_2^2\!+\!c_2^2d_2^2f_2^2+c_3^2d_3^2e_3^2+c_3^2d_3^2f_3^2).
\end{eqnarray}
From (\ref{rho^2}), (\ref{rhochi2}) and (\ref{chi2}), we obtain
\begin{eqnarray}
||\rho-\chi||^2={\rm Tr}(\rho^2)-2{\rm Tr}(\rho\chi)+{\rm Tr}(\chi^2)\nonumber\\
=\frac{1}{8}\{c_1^2+c_2^2+c_3^2-\frac{1}{2}[c_1^2d_1^2(e_1^2+f_1^2)+c_2^2d_2^2(e_2^2+f_2^2)+c_3^2d_3^2(e_3^2+f_3^2)]\}.\nonumber
\end{eqnarray}
It can be directly checked that $d_1^2+d_2^2+d_3^2=1$, $e_1^2+e_2^2+e_3^2=1$ and $f_1^2+f_2^2+f_3^2=1.$
Set
\begin{eqnarray}\label{c}
c:=\max\{|c_1|,|c_2|,|c_3|\}.
\end{eqnarray}
Then
\begin{eqnarray}\label{c3}
\frac{1}{2}[c_1^2d_1^2(e_1^2+f_1^2)+c_2^2d_2^2(e_2^2+f_2^2)+c_3^2d_3^2(e_3^2+f_3^2)]
\nonumber\\
\leq \frac{1}{2}|c|^2[d_1^2(e_1^2+f_1^2)+d_2^2(e_2^2+f_2^2)+d_3^2(e_3^2+f_3^2)]
=c^2,
\end{eqnarray}
in which this equality can be easily obtained by appropriate choice of $t_A$, $t_{B^0}$, $t_{B^1}$, $y_{Aj}$, $y_{B^0j}$ and $y_{B^1j}$. In particular,
the equality in (\ref{c3}) holds for the following cases:
(1) If $c=|c_1|$, then $|d_1|=|e_1|=|f_1|=1,d_2=d_3=e_2=e_3=f_2=f_3=0$. For example, $|t_A|=|y_{A2}|=|t_{B^0}|=|y_{B^02}|=|t_{B^1}|=|y_{B^12}|=\frac{1}{\surd2}$ and $y_{A1}=y_{A3}=y_{B^01}=y_{B^03}=y_{B^11}=y_{B^13}=0.$
(2) If $c=|c_2|$, then $|d_2|=|e_2|=|f_2|=1,d_1=d_3=e_1=e_3=f_1=f_3=0$. For instance, $|t_A|=|y_{A1}|=|t_{B^0}|=|y_{B^01}|=|t_{B^1}|=|y_{B^11}|=\frac{1}{\surd2}$ and $y_{A2}=y_{A3}=y_{B^02}=y_{B^03}=y_{B^12}=y_{B^13}=0.$
(3) If $c=|c_3|,$ then $|d_3|=|e_3|=|f_3|=1,d_1=d_2=e_1=e_2=f_1=f_2=0,$ e.g., $y_{A1}=y_{A2}=y_{B^01}=y_{B^02}=y_{B^11}=y_{B^12}=0.$
Therefore, we have
\begin{eqnarray}
D^{(3)}_G(\rho)=\frac{1}{8}(c_1^2+c_2^2+c_3^2-c^2).
\end{eqnarray}

Now we consider the family of four-qubit states, associated with systems $A$, $B$, $C$, and $D$,
\begin{equation}\label{rho4}
\rho=\frac{1}{16}(I+\sum\limits_{j=1}^3c_j\sigma_j\otimes\sigma_j\otimes\sigma_j\otimes\sigma_j).
\end{equation}
One has ${\rm Tr}(\rho^2)=\frac{1}{16}(1+c_1^2+c_2^2+c_3^2)$.

With respect to the local measurement on the subsystem $A$, we obtain
\begin{eqnarray}
\rho_0\!=\!\frac{1}{8}V_{A}\Pi_0V_{A}^\dagger\!\otimes\!(I\!\otimes\! I\!\otimes\! I\!+\!c_1g_{1}\sigma_1\!\otimes\!\sigma_1\!\otimes\!\sigma_1
\!+\!c_2g_{2}\sigma_2\!\otimes\!\sigma_2\!\otimes\!\sigma_2\!+\!c_3g_{3}\sigma_3\!
\otimes\!\sigma_3\!\otimes\!\sigma_3),\nonumber
\end{eqnarray}
\begin{eqnarray}
\rho_1\!=\!\frac{1}{8}V_{A}\Pi_1V_{A}^\dagger\!\otimes\!(I\!\otimes\! I\!\otimes\! I\!-\!c_1g_{1}\sigma_1\!\otimes\!\sigma_1\!\otimes\!\sigma_1
\!-\!c_2g_{2}\sigma_2\!\otimes\!\sigma_2\!\otimes\!\sigma_2\!-\!c_3g_{3}\sigma_3\!
\otimes\!\sigma_3\!\otimes\!\sigma_3),\nonumber
\end{eqnarray}
where
\begin{eqnarray}
& g_{1}=2(-t_Ay_{A2}+y_{A1}y_{A3}),\nonumber \\
& g_{2}=2(t_Ay_{A1}+y_{A2}y_{A3}),\nonumber \\
& g_{3}=t_A^2-y_{A1}^2-y_{A2}^2+y_{A3}^2.\nonumber
\end{eqnarray}

After the subsequent measurement on $B$, we get
\begin{eqnarray}
\rho_{00}\!=\!\frac{1}{4}V_{A}\Pi_0V_{A}^\dagger\!\otimes\! V_{{B}^0}\Pi_0V_{{B}^0}^\dagger\!\otimes\!(I\!\otimes \! I\!+\!c_1g_{1}h_1\sigma_1\!\otimes\!\sigma_1\!+\!c_2g_{2}h_2\sigma_2\!\otimes\!
\sigma_2\!+\!c_3g_{3}h_3\sigma_3\!\otimes\!\sigma_3),\nonumber\\
\rho_{01}\!=\!\frac{1}{4}V_{A}\Pi_0V_{A}^\dagger\!\otimes \! V_{{B}^0}\Pi_1V_{{B}^0}^\dagger\!\otimes\!(I\!\otimes \! I\!-\!c_1g_{1}h_1\sigma_1\!\otimes\!\sigma_1\!-\!c_2g_{2}h_2\sigma_2\!\otimes\!
\sigma_2\!-\!c_3g_{3}h_3\sigma_3\!\otimes\!\sigma_3),\nonumber\\
\rho_{10}\!=\!\frac{1}{4}V_{A}\Pi_1V_{A}^\dagger\!\otimes\! V_{{B}^1}\Pi_0V_{{B}^1}^\dagger\!\otimes\!(I\!\otimes\! I\!-\!c_1g_{1}m_1\sigma_1\!\otimes\!\sigma_1\!-\!c_2g_{2}m_2\sigma_2\!\otimes\!
\sigma_2\!-\!c_3g_{3}m_3\sigma_3\!\otimes\!\sigma_3),\nonumber\\
\rho_{11}\!=\!\frac{1}{4}V_{A}\Pi_1V_{A}^\dagger\!\otimes \! V_{{B}^1}\Pi_1V_{{B}^1}^\dagger\!\otimes\!(I\!\otimes \! I\!+\!c_1g_{1}m_1\sigma_1\!\otimes\!\sigma_1\!+\!c_2g_{2}m_2\sigma_2\!\otimes\!
\sigma_2\!+\!c_3g_{3}m_3\sigma_3\!\otimes\!\sigma_3),\nonumber
\end{eqnarray}
where
\begin{eqnarray}
& h_{1}=2(-t_{{B}^0}y_{{B}^02}+y_{{B}^01}y_{{B}^03}),\nonumber \\
& h_{2}=2(t_{{B}^0}y_{{B}^01}+y_{{B}^02}y_{{B}^03}),\nonumber\\
& h_{3}=t_{{B}^0}^2-y_{{B}^01}^2-y_{{B}^02}^2+y_{{B}^03}^2,\nonumber
\end{eqnarray}
and
\begin{eqnarray}
& m_{1}=2(-t_{{B}^1}y_{{B}^12}+y_{{B}^11}y_{{B}^13}),\nonumber \\
& m_{2}=2(t_{{B}^1}y_{{B}^11}+y_{{B}^12}y_{{B}^13}),\nonumber \\
& m_{3}=t_{{B}^1}^2-y_{{B}^11}^2-y_{{B}^12}^2+y_{{B}^13}^2.\nonumber
\end{eqnarray}

Based on the measurement outcomes from $A$ and $B$, the measurement on the subsystem $C$ give rise to
\begin{eqnarray}
\rho_{000}=\frac{1}{2}V_{A}\Pi_0V_{A}^\dagger\otimes V_{{B}^0}\Pi_0V_{{B}^0}^\dagger\otimes V_{{C}^{00}}\Pi_0V_{{C}^{00}}^\dagger\otimes(I+c_1g_1h_1n_1\sigma_1+c_2g_2h_2n_2\sigma_2+c_3g_3h_3n_3\sigma_3),\nonumber\\
\rho_{001}=\frac{1}{2}V_{A}\Pi_0V_{A}^\dagger\otimes V_{{B}^0}\Pi_0V_{{B}^0}^\dagger\otimes V_{{C}^{00}}\Pi_1V_{{C}^{00}}^\dagger\otimes(I-c_1g_1h_1n_1\sigma_1-c_2g_2h_2n_2\sigma_2-c_3g_3h_3n_3\sigma_3),\nonumber\\
\rho_{010}=\frac{1}{2}V_{A}\Pi_0V_{A}^\dagger\otimes V_{{B}^0}\Pi_1V_{{B}^0}^\dagger\otimes V_{{C}^{01}}\Pi_0V_{{C}^{01}}^\dagger\otimes(I-c_1g_1h_1o_1\sigma_1-c_2g_2h_2o_2\sigma_2-c_3g_3h_3o_3\sigma_3),\nonumber\\
\rho_{011}=\frac{1}{2}V_{A}\Pi_0V_{A}^\dagger\otimes V_{{B}^0}\Pi_1V_{{B}^0}^\dagger\otimes V_{{C}^{01}}\Pi_1V_{{C}^{01}}^\dagger\otimes(I+c_1g_1h_1o_1\sigma_1+c_2g_2h_2o_2\sigma_2+c_3g_3h_3o_3\sigma_3),\nonumber\\
\rho_{100}=\frac{1}{2}V_{A}\Pi_1V_{A}^\dagger\otimes V_{{B}^1}\Pi_0V_{{B}^1}^\dagger\otimes V_{{C}^{10}}\Pi_0V_{{C}^{10}}^\dagger\otimes(I-c_1g_1m_1r_1\sigma_1-c_2g_2m_2r_2\sigma_2-c_3g_3m_3r_3\sigma_3),\nonumber\\
\rho_{101}=\frac{1}{2}V_{A}\Pi_1V_{A}^\dagger\otimes V_{{B}^1}\Pi_0V_{{B}^1}^\dagger\otimes V_{{C}^{10}}\Pi_1V_{{C}^{10}}^\dagger\otimes(I+c_1g_1m_1r_1\sigma_1+c_2g_2m_2r_2\sigma_2+c_3g_3m_3r_3\sigma_3),\nonumber\\
\rho_{110}=\frac{1}{2}V_{A}\Pi_1V_{A}^\dagger\otimes V_{{B}^1}\Pi_1V_{{B}^1}^\dagger\otimes V_{{C}^{11}}\Pi_0V_{{C}^{11}}^\dagger\otimes(I+c_1g_1m_1s_1\sigma_1+c_2g_2m_2s_2\sigma_2+c_3g_3m_3s_3\sigma_3),\nonumber\\
\rho_{111}=\frac{1}{2}V_{A}\Pi_1V_{A}^\dagger\otimes V_{{B}^1}\Pi_1V_{{B}^1}^\dagger\otimes V_{{C}^{11}}\Pi_1V_{{C}^{11}}^\dagger\otimes(I-c_1g_1m_1s_1\sigma_1-c_2g_2m_2s_2\sigma_2-c_3g_3m_3s_3\sigma_3),\nonumber
\end{eqnarray}
where the index $u$ ($v$) in the unitary $\{V_{C^{uv}}:~u=0,1;\,v=0,1\}$ corresponds to the outcome of the measurement on $A$ ($B$).

The post measurement state is given as $\chi=p_{000}\rho_{000}+p_{001}\rho_{001}+p_{010}\rho_{010}+p_{011}\rho_{011}
+p_{100}\rho_{100}+p_{101}\rho_{101}+p_{110}\rho_{110}+p_{111}\rho_{111}$.
Therefore, we have
\begin{eqnarray}
-2{\rm Tr}(\rho\chi)&=&-\frac{1}{8}[{\rm Tr}(I\chi)+{\rm Tr}(\sum\limits_{j=1}^3c_j\sigma_j\otimes\sigma_j\otimes\sigma_j\otimes\sigma_j\chi)]\nonumber\\
&=&\!-\!\frac{1}{8}[1\!+\!\frac{1}{4}(c_1^2g_1^2h_1^2n_1^2\!+\!c_1^2g_1^2h_1^2o_1^2\!+\!c_1^2g_1^2m_1^2r_1^2\!+\!c_1^2g_1^2m_1^2s_1^2
\!+\!c_2^2g_2^2h_2^2n_2^2\!+\!c_2^2g_2^2h_2^2o_2^2\nonumber\\
&&+c_2^2g_2^2m_2^2r_2^2+c_2^2g_2^2m_2^2s_2^2+c_3^2g_3^2h_3^2n_3^2+c_3^2g_3^2h_3^2o_3^2+c_3^2g_3^2m_3^2r_3^2+c_3^2g_3^2m_3^2s_3^2)],\nonumber
\end{eqnarray}
\begin{eqnarray}
{\rm Tr}(\chi^2)=&\frac{1}{64}(4+c_1^2g_1^2h_1^2n_1^2+c_1^2g_1^2h_1^2o_1^2+c_1^2g_1^2m_1^2r_1^2+c_1^2g_1^2m_1^2s_1^2
+c_2^2g_2^2h_2^2n_2^2+c_2^2g_2^2h_2^2o_2^2\nonumber\\
&+c_2^2g_2^2m_2^2r_2^2+c_2^2g_2^2m_2^2s_2^2+c_3^2g_3^2h_3^2n_3^2+c_3^2g_3^2h_3^2o_3^2+c_3^2g_3^2m_3^2r_3^2+c_3^2g_3^2m_3^2s_3^2)\nonumber
\end{eqnarray}
and
\begin{eqnarray}
||\rho-\chi||^2&=&{\rm Tr}(\rho^2)-2{\rm Tr}(\rho\chi)+{\rm Tr}(\chi^2)\nonumber\\
&=&\frac{1}{16}[c_1^2+c_2^2+c_3^2-\frac{1}{4}(c_1^2g_1^2h_1^2n_1^2+c_1^2g_1^2h_1^2o_1^2+c_1^2g_1^2m_1^2r_1^2
\nonumber\\
&&+c_1^2g_1^2m_1^2s_1^2
+c_2^2g_2^2h_2^2n_2^2+c_2^2g_2^2h_2^2o_2^2
+c_2^2g_2^2m_2^2r_2^2+c_2^2g_2^2m_2^2s_2^2\nonumber\\
&&+c_3^2g_3^2h_3^2n_3^2+c_3^2g_3^2h_3^2o_3^2+c_3^2g_3^2m_3^2r_3^2+c_3^2g_3^2m_3^2s_3^2)].\nonumber
\end{eqnarray}

It can be directly verified that $g_1^2+g_2^2+g_3^2=1,$ $h_1^2+h_2^2+h_3^2=1,$ $m_1^2+m_2^2+m_3^2=1,$ $n_1^2+n_2^2+n_3^2=1,$  $o_1^2+o_2^2+o_3^2=1,$ $r_1^2+r_2^2+r_3^2=1$ and $s_1^2+s_2^2+s_3^2=1.$ Then
$\frac{1}{4}(c_1^2g_1^2h_1^2n_1^2+c_1^2g_1^2h_1^2o_1^2+c_1^2g_1^2m_1^2r_1^2+c_1^2g_1^2m_1^2s_1^2
+c_2^2g_2^2h_2^2n_2^2+c_2^2g_2^2h_2^2o_2^2++c_2^2g_2^2m_2^2r_2^2+c_2^2g_2^2m_2^2s_2^2
+c_3^2g_3^2h_3^2n_3^2+c_3^2g_3^2h_3^2o_3^2+c_3^2g_3^2m_3^2r_3^2+c_3^2g_3^2m_3^2s_3^2)]\leq c^2.$
Finally, the geometric discord for the four-qubit state is give by
\begin{eqnarray}
D^{(4)}_G(\rho)=\frac{1}{16}(c_1^2+c_2^2+c_3^2-c^2).
\end{eqnarray}

For the case of general multi-qubit state, we have

\begin{theorem}
For the family of $N$-qubit ($N\geq2$) states (\ref{rho}), the geometric discord is given by
\begin{eqnarray}\label{dngrho}
D^{(N)}_G(\rho)=\frac{1}{2^N}(c_1^2+c_2^2+c_3^2-c^2),
\end{eqnarray}
where $c=\max\{|c_1|,|c_2|,|c_3|\}$.
\end{theorem}

{\sf [Proof]}
For the (\ref{rho}), one has ${\rm Tr}(\rho^2)=\frac{1}{2^N}(1+c_1^2+c_2^2+c_3^2)$.
After $N-1$ measurements, we obtain
 \begin{eqnarray}
&p_1\rho_1=\frac{1}{2^N}V_{A_1}\Pi_0V_{A_1}^\dagger\otimes \cdots\otimes V_{A_{N-1}}\Pi_0V_{A_{N-1}}^\dagger\otimes q_1,\nonumber\\
&p_2\rho_2=\frac{1}{2^N}V_{A_1}\Pi_0V_{A_1}^\dagger\otimes \cdots\otimes V_{A_{N-1}}\Pi_1V_{A_{N-1}}^\dagger\otimes q_2,\nonumber\\
&\cdots\nonumber\\
&p_{2^{N-1}}\rho_{2^{N-1}}=\frac{1}{2^N}V_{A_1}\Pi_1V_{A_1}^\dagger\otimes \cdots\otimes V_{A_{N-1}}\Pi_1V_{A_{N-1}}^\dagger\otimes q_{2^{N-1}},\nonumber
\end{eqnarray}
where $q_k$ are the quantum states in $A_N$.
The state $\chi$ is given as $\chi=p_1\rho_1+p_2\rho_2+\cdots+p_{2^{N-1}}\rho_{2^{N-1}}.$
Therefore,
\begin{eqnarray}
&-2{\rm Tr}(\rho\chi)=-\frac{1}{2^{N-1}}[{\rm Tr}
(I\chi)+{\rm Tr}(\sum\limits_{j=1}^3c_j\sigma_j\otimes...\otimes\sigma_j\chi)],\nonumber\\
&{\rm Tr}(\chi^2)=\frac{1}{2^{2N}}({\rm Tr}{q_1^2}+{\rm Tr}{q_2^2}+...+{\rm Tr}{q_{2^{N-1}}^2}).\nonumber
\end{eqnarray}
By (\ref{c}) we can easily evaluate that
\begin{eqnarray}
\min(-2{\rm Tr}(\rho\chi))=-\frac{1}{2^{N-1}}(1+c^2),\nonumber\\
\min(2{\rm Tr}(\chi^2))=\frac{1}{2^N}(1+c^2).\nonumber
\end{eqnarray}
Hence, the geometric discord for (\ref{rho}) is given by $D^{(N)}_G(\rho)=\frac{1}{2^N}(c_1^2+c_2^2+c_3^2-c^2)$. $\Box$

\begin{figure*}[h]
\begin{center}
\includegraphics[width=10.9cm]{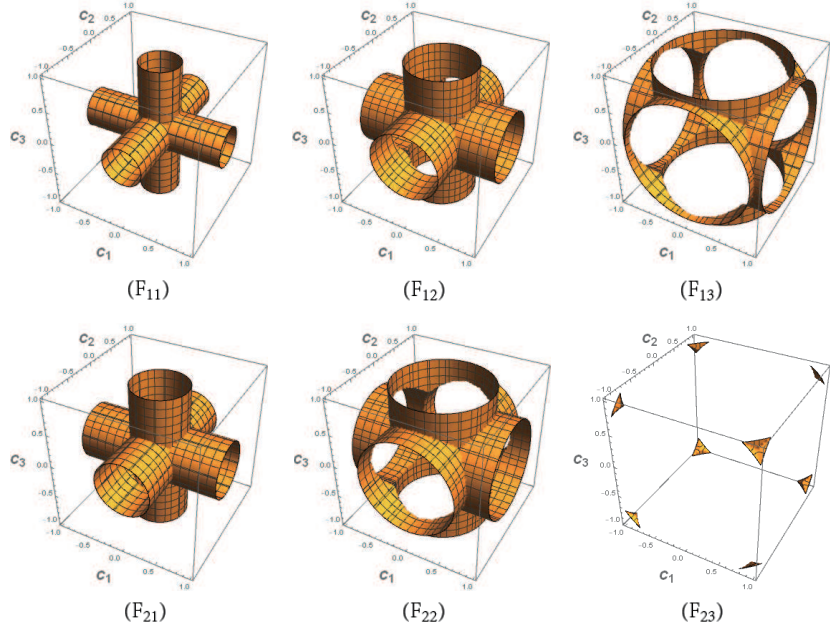}
\caption{Level surfaces of constant geometric discord.
$N=3$ for figures $(F_{11})$, $(F_{12})$ and $(F_{13})$ with $D_G(\rho)=0.01,~0.03$ and $0.1$, respectively.
$N=4$ for figures $(F_{21})$, $(F_{22})$ and $(F_{23})$ with $D_G(\rho)=0.01,~0.03$ and $0.1$, respectively.}
\end{center}
\label{Fig:1}
\end{figure*}

Yao \emph{et.al.} \cite{Yao} has already compared discord to geometric discord when $N=2$.
They obtained that the level surfaces of geometric discord are consisted of three identical intersecting "cylinders" rather than irregular "tubes".
Figure 1 shows the level surfaces of geometric discord for $N=3$ and $4$.
For small discord, the level surfaces are centrally symmetric, consisting of three intersecting "cylinders" along the three coordinate axes.
For larger discord, these intersecting tubes keep expanding.
And as shown in $(F_{23})$, it finally expands until only a few vertices remained.
Compared with level surfaces of quantum discord depicted in \cite{Li}, all of these phenomena are very similar to discord.
Since we have discovered in \cite{Li} that the quantum discord of this family of states can be classified into three categories,
and it is found that only the coefficient of geometric discord will change for states with different number of qubits in this article.
We obtain that for this family of states, when $N$ is fixed, geometric discord can reflect the change in discord to some extent.

\section{Geometric discord under single qubit noise}
As is well known, the geometric discord for some states may change  suddenly under some decoherence channels \cite{Maziero,Jia,Yao,Montealegre,Hu,Yan}.
It would be interesting to know if such phenomena exist when only one of the qubits subjects to a noisy environment.
We first consider the Bell-diagonal states under the phase flip channel $\varepsilon(\cdot)$, with the Kraus operators
$\Gamma_0=$ diag$(1,\gamma)\otimes I ,$ $\Gamma_1=$ diag$(0,\sqrt{1-\gamma^2})\otimes I ,$ where $\gamma=e^{-\frac{\tau t}{2}}$ and $\tau$ denotes transversal decay rate.
One gets
\begin{eqnarray}
\varepsilon(\rho)=\Gamma_0\rho\Gamma_0^\dag+\Gamma_1\rho\Gamma_1^\dag=\frac{1}{4}(I+\gamma c_1\sigma_1\otimes\sigma_1
+\gamma c_2\sigma_2\otimes\sigma_2+c_3\sigma_3\otimes\sigma_3).\nonumber
\end{eqnarray}
\begin{eqnarray}\label{d2g}
D^{(2)}_G(\varepsilon(\rho))=\frac{1}{4}[\gamma^2(c_1^2+c_2^2)+c_3^2-\max\{(\gamma c_1)^2,(\gamma c_2)^2,c_3^2\}].
\end{eqnarray}
If $|c_3|\geq \max\{|c_1|,|c_2|\}$, the geometric discord $D^{(2)}_G(\varepsilon(\rho))$
equals to $\frac{\gamma^2}{4}(c_1^2+c_2^2)$, which decays monotonically.
If $\max\{|c_1|,|c_2|\}\geq|c_3|$ and $|c_3|\neq 0$, the geometric discord $D^{(2)}_G(\varepsilon(\rho))$
has a sudden change at 
$t_0=-\frac{2}{\tau}ln\frac{\max\{|c_1|,|c_2|\}}{|c_3|}.$
The dynamic behavior of the geometric discord for Bell-diagonal states with different $\{c_i\} $ is depicted in Figure 2(a).
It is shown that sudden change of geometric discord also occurs when the phase noise acts only on one of the qubits.
Compared with Jia \emph{et.al.}'s conclusions \cite{Jia}, we discover that geometric discord can reflect quantum discord changes in the case of local phase inversion.

\begin{figure}[htbp]
\centering{\includegraphics[width=12cm]{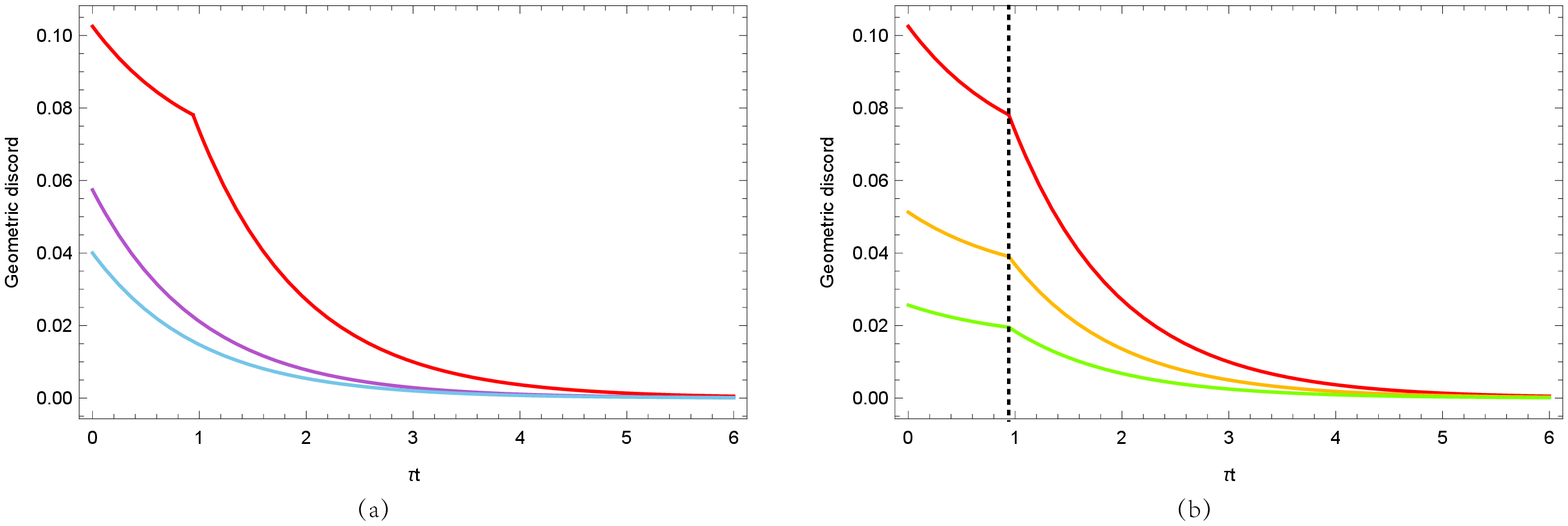}}
\caption{Geometric discord of two-qubit, three-qubit and four-qubit states under local phase flip channels.
(a) Geometric discord of two-qubit states under local phase flip channels. (a1) $c_1=\frac{4}{5}$, $c_2=\frac{c_1}{2}$, $c_3=\frac{1}{2}$ (red line); (a2) $c_1=\frac{3}{7}$, $c_2=\frac{c_1}{2}$ $c_3=\frac{4}{5}$ (purple line); (a3) $c_1=\frac{4}{5}$, $c_2=\frac{c_1}{2}$, $c_3=0$ (blue line).
(b) The multipartite geometric discord when $c_1=\frac{4}{5}$, $c_2=\frac{c_1}{2}$, $c_3=\frac{1}{2}$. (b1) $D^{(2)}_G(\varepsilon(\rho)) $(red line); (b2) $D^{(3)}_G(\varepsilon(\rho))$ (orange line); (b3) $D^{(4)}_G(\varepsilon(\rho))$ (green line).}
\label{transition}
\end{figure}

Now we consider three-qubit states (\ref{rho3}) under the phase flip channel for single qubit, with the Kraus operators
$\Gamma_0=$ diag$(1,\gamma)\otimes I\otimes I$ and $\Gamma_1=$ diag$(0,\sqrt{1-\gamma^2})\otimes I\otimes I$. We have
\begin{eqnarray}
\varepsilon(\rho)=\frac{1}{8}(I+\gamma c_1\sigma_1\otimes\sigma_1\otimes\sigma_1
+\gamma c_2\sigma_2\otimes\sigma_2\otimes\sigma_2+c_3\sigma_3\otimes\sigma_3\otimes\sigma_3),\nonumber
\end{eqnarray}
\begin{eqnarray}\label{d3g}
D^{(3)}_G(\varepsilon(\rho))=\frac{1}{8}[\gamma^2(c_1^2+c_2^2)+c_3^2-\max\{(\gamma c_1)^2,(\gamma c_2)^2,c_3^2\}].
\end{eqnarray}

Similarly, for the four-qubit states (\ref{rho4}) under the operators of phase noise acting on the first qubit, wiht
$\Gamma_0=$ diag$(1,\gamma)\otimes I\otimes I\otimes I$ and $\Gamma_1=$ diag$(0,\sqrt{1-\gamma^2})\otimes I\otimes I\otimes I$, we obtain
\begin{eqnarray}
\varepsilon(\rho)=\frac{1}{16}(I+\gamma c_1\sigma_1\otimes\sigma_1\otimes\sigma_1
+\gamma c_2\sigma_2\otimes\sigma_2\otimes\sigma_2+c_3\sigma_3\otimes\sigma_3\otimes\sigma_3),\nonumber
\end{eqnarray}
\begin{eqnarray}\label{d4g}
D^{(4)}_G(\varepsilon(\rho))=\frac{1}{16}[\gamma^2(c_1^2+c_2^2)+c_3^2-\max\{(\gamma c_1)^2,(\gamma c_2)^2,c_3^2\}].
\end{eqnarray}

Fig. 2(b) shows the dynamical behavior of the multipartite geometric discord $D^{(2)}_G(\varepsilon(\rho))$, $D^{(3)}_G(\varepsilon(\rho))$, and $D^{(4)}_G(\varepsilon(\rho))$ where the sudden change exists when $\max\{|c_1|,|c_2|\}\geq|c_3|$ and $|c_3|\neq 0.$
Moreover, the sudden change occurs at $t_0=-\frac{2}{\tau}ln\frac{\max\{|c_1|,|c_2|\}}{|c_3|}$ .
Therefore, for the same $\{c_i\}$, they make sudden changes at the same time.

\section{conclusion}
The bipartite quantum discord had been introduced by Ollivier and Zurek \cite{Ollivier} in 2001. Recently, Radhakrishnan et. al provide the multipartite quantum discord \cite{Radhakrishnan}.
According to bipartite geometric discord and multipartite quantum discord, we have introduced the geometric discord for multipartite states,
with each measurement depends conditionally on the previous measurement outcomes.
We have explicitly derived the geometric discord for $N$-qubit states (1).
Furthermore, we have shown that the sudden change of the multi-qubit geometric discord also appears when the phase noise acts only on one of the qubits.
Our results may highlight further investigations on multipartite geometric discord and the applications in quantum information processing.

\bigskip
\noindent{\bf Acknowledgments}\, \, This work is supported by NSFC under numbers 11765016 and 12075159, the
GJJ170444, Beijing Natural Science Foundation (Z190005), Academy for Multidisciplinary Studies, Capital Normal University,
and Shenzhen Institute for Quantum Science and Engineering, Southern University of Science and Technology (No. SIQSE202001).

\end{document}